\documentclass[onecolumn]{aastex631}

\usepackage{amsmath,amssymb,amsfonts}%
\usepackage{amsthm}%
\usepackage{mathrsfs}%
\usepackage{graphicx}
\usepackage{enumitem}

\begin{document}

\title{New Approach to Superflare Energy Determination}

\author[0000-0002-5778-2600]{P. Heinzel}
\affiliation{University of Wrocław, Centre of Scientific Excellence - Solar and Stellar Activity, Kopernika 11, 51-622 Wrocław, Poland}
\affiliation{Astronomical Institute, Academy of Sciences of the Czech Republic, 25165 Ond\v{r}ejov, Czech Republic}

\author[0000-0003-1853-2809]{R. Falewicz}
\affiliation{Astronomical Institute, University of Wrocław, Kopernika 11, 51-622 Wrocław, Poland}
\affiliation{University of Wrocław, Centre of Scientific Excellence - Solar and Stellar Activity, Kopernika 11, 51-622 Wrocław, Poland}

\author[0000-0003-1419-2835]{K. Bicz}
\affiliation{Astronomical Institute, University of Wrocław, Kopernika 11, 51-622 Wrocław, Poland}

\author[0000-0001-8474-7694]{P. Pre\'s}
\affiliation{Astronomical Institute, University of Wrocław, Kopernika 11, 51-622 Wrocław, Poland}



\begin{abstract}

\noindent
We present a new method for estimating the total energy radiated by stellar flares in broad-band continua, which assumes a constant emitting area but incorporates a time-dependent temperature evolution. This physically motivated approach offers an alternative to the commonly used method that assumes a fixed flare temperature about of 10 000\,K and variable area. By allowing the temperature to vary over time while keeping the emitting area constant, our method captures more realistic flare behaviour. This time-dependent treatment of the flare temperature is supported by numerous solar observations, numerical simulations, and multiwavelength studies of active stars.
We demonstrate that using peak flare temperatures estimated from a semi-empirical model grid, rather than assuming an ad-hoc flare temperature value, improves the accuracy of total energy estimates. Although the most precise results still require a multi-band photometry/spectroscopy or independently constrained flare temperatures, our method offers a practical and scalable solution for single-band observations. It is particularly well suited for main-sequence stars of spectral types K4 and later with known effective temperatures. Finally we discuss how the flare continuum behaves under varying chromospheric conditions.
Our method improves flare energy estimates by incorporating a physically relevant time-dependent temperature evolution and empirically derived peak temperatures, rather than assuming a constant 10 000\,K value. This modification reduces systematic errors that can reach factors up to ten as compared to previous estimates. We proved this on a sample of 50,320 TESS flares.

\end{abstract}

\section{Introduction}\label{sec:intro}

\noindent The detection of enormous number of stellar flares on cool stars by missions such as {\it Kepler/K2} \citep{Keplercite, Howell_2014}, and TESS (Transiting Exoplanet Survey Satellite) \citep{Ricker_2014} has enabled large-scale studies of flare energetics. These flares, which occur on M-, K-, and G-type stars, exhibit a wide range of amplitudes and durations, from typical flares lasting minutes to hours to rare, long-duration events extending over a day \citep{Gunther_2020, Pietras22, Bicz_20242}. Particularly energetic events—so-called superflares with energies exceeding 10$^{33}$ erg—have been observed, with the most extreme cases reaching or surpassing 10$^{36}$ erg \citep{Schaefer_2000, Maehara12, Shibayama13, Gunther_2020, Pietras22}.

Traditionally, the total energy emitted during such flares is estimated by modeling the continuum emission as blackbody radiation with a constant chromospheric temperature of about 10$\,$000$\,$K. The flare area is inferred from light-curve amplitudes and assumed to vary in time, with the total energy determined by integrating over the flare duration \citep{Shibayama13}. This approach has formed the basis of numerous statistical analyses, including recent work based on TESS data \citep{Pietras22, Seli_2025}.
Nevertheless, it is based on assumptions that seem to contradict both observational data and the results of numerical simulations.
In reality, the flare temperature evolves throughout the event, typically peaking during the impulsive phase and declining during the gradual phase as the chromosphere receives diminishing energy input

\citep{Kowalski_2013, Howard_2020, Bicz_2025}. At the same time, the flaring area remains relatively stable over much of the flare’s duration \citep{Bicz_2025}.

Ignoring the temperature evolution can lead to significant systematic errors—potentially of several hundred percents in estimates of the flare’s bolometric energy, particularly when derived from single-band photometric data. Accurate flare-energy measurements are essential for understanding the influence of stellar activity on exoplanetary environments, including space weather conditions, atmospheric erosion, ozone depletion \citep{Tilley_2019}, and UV-driven prebiotic chemistry \citep{Rimmer_2018, Gunther_2020}.

Previous works often assumed a constant flare temperature of 10 000\,K \citep{1980ApJ...239L..27M,1992ApJS...78..565H,2015SoPh..290.3663K}, ignoring observational evidence of strong thermal evolution during flares. Because bolometric luminosity strongly scales with temperature, this simplification introduces systematic biases, especially in single-band photometry. Our method addresses this issue by allowing temperature to vary over time and by estimating peak temperature from semi-empirical grids. This is crucial because the canonical 10 000\,K value is reached only briefly (5--19\% of flare duration) or not at all, and real peak temperatures can differ by thousands of Kelvin, leading to errors up to an order of magnitude in energy estimates.

In this study, we propose a revised method for estimating superflare energies. Our approach allows the chromospheric flare temperature to evolve over time, following the shape of the light curve, while the flare area is assumed to remain approximately constant—an assumption supported by multi-band photometric observations \citep{Bicz_2025} and confirmed by recent high-resolution solar observations (see below). The corresponding energy calculation formulas are presented in Section 2.
In Section 3, we apply the method to a superflare observed by TESS and compare the results with those obtained using the standard approach described in \citet{Shibayama13}, hereafter referred to as Method I.
Section 4 provides statistical insights from a larger sample of TESS flare events and highlights the relevance of our method for single-band photometric missions, such as {\it Kepler}, TESS or PLATO. Finally, in Section 5, we summarize our findings and discuss the applicability and limitations of the proposed energy estimation method for stellar flares.

\section{Superflare energy determination}

\subsection{Method I Described in \cite{Shibayama13}}
\noindent Estimating the total (bolometric) continuum emission requires assumptions about the thermal evolution of the flaring region. Without multi-band or spectral data, we are forced to simplify the calculations.
The assumption that the flare continuum follows a Planckian spectral distribution is reasonable only for strong flares, particularly superflares, where we expect a high optical thickness in the Paschen continuum (proportional to the expected high densities squared), leading to thermalization of the continuum towards blackbody radiation (see, e.g., \cite{Heinzel2024}). For weaker flares, such as those observed on the Sun, continuum formation is more complex, as discussed by \cite{Simoes24} and \cite{Heinzel2024}.

The method presented in \cite{Shibayama13} (Method I) is based on the key assumption that the temperature of the flaring plasma remains constant over time, typically set at 10$\,$000$\,$K or close to this value. Such an assumption allows for the determination of the temporal evolution of the flaring area, $A_{\rm flare}$, which is taken to be proportional to the flare amplitude observed in the light curve.
Furthermore, the original method assumes that the flare area is much smaller than the total stellar surface area, i.e., $A_{\rm flare} \ll A_{\rm star}$, where $A_{\rm star}=4 \pi R^2_{\rm star}$.
Once the time-dependent flare area is determined
the total flare energy can be estimated as the integral over flare
duration:
\begin{equation}
    E_{\rm{flare}} = \sigma_{\rm SB} T^4_{\rm flare }\int\limits_{\rm{flare}}A_{\rm flare}(t) dt\, ,
\end{equation}
where $T_{\rm flare} = 10^4$ K and $\sigma_{\rm SB}$ is the
Stefan-Boltzmann constant
(for details see \citet{Shibayama13}). This technique has shown that strong flares detected by {\it Kepler} and \emph{TESS} can reach energies of up to 10$^{36}\,$erg or more—at least four orders of magnitude greater than the strongest solar flares. Given that many flare observations are limited to a single photometric band, developing simplified yet still accurate energy estimation techniques is essential.

\subsection{Method IIa Using a Constant Flare Area}\label{subs:constanta}

\noindent Here, we propose a simple yet physically more realistic modification of Method I. Instead of assuming a constant flare temperature throughout the duration of the event, we fix the flare area and allow the temperature to evolve over time. Introducing a time-dependent temperature is more physically justified, as demonstrated by numerous solar observations (e.g. recently
\citet{2024A&A...690A.254G,2025A&A...697A.103F}),
often supported by numerical simulations \citep{2023A&A...673A.150C}, as well as multiwavelength observations of flare-active stars \citep{Bicz_2025}.

For example, in contrast to the method described above, the assumption of a (quasi-)constant flare emission area reflects the idea that the spatial extent of the active region in which the flare occurs is regulated by a relatively stable magnetic field topology.
Within this framework, assuming a Planckian spectrum, brightness variations during a flare are attributed primarily to changes in temperature rather than to variations in the emitting area.
Solar observations support such behavior, especially regarding the stability of the magnetic field topology during flares \citep{Zhao_2014, Gupta_2024}. Both studies provide strong support for the assumption of magnetic field topology stability during flaring events. In the first work, the authors analyze a series of 3D magnetic field models for active regions during major flares and find that the magnetic topology, specifically the height and structure of the field above the active region, does not undergo significant changes over the flare duration.
They conclude that the field structure remains relatively stable, and that the differences in flare behavior arise primarily from critical conditions and the level of helicity. The second study, a case analysis of a specific active region, focuses on two major flares. Based on magnetic field reconstruction from SDO/HMI (Solar Dynamics Observatory / Helioseismic and Magnetic Imager) data, the authors demonstrate that the key topological structure—the so-called hyperbolic flux tube (HFT)—remains largely unchanged during the course of strong solar flares (M6.6 and X2.2), indicating its stability on hourly timescales. These studies form the basis for the assumptions presented above.

It can also be argued that flare area may appear to expand due to gradual magnetic reconnection and the rising of flare loops (see, e.g., \citet{Dudik_2014}). However, our assumption of a fixed area represents a reasonable compromise. On the Sun, for instance, as flare ribbons gradually fade, we observe their return to the quiet atmosphere characterized by the stellar temperature $T_{\rm star}$.

Although the flare area may visually appear to shrink as the event declines, we interpret it more accurately as the region occupied by the footpoints of flare arcades. During flare evolution, these arcades tend to remain geometrically stable, and we often observe coronal loops well into the later stages of flares. However, the emission from their footprints, i.e., the flare ribbons, decreases as the flare heating becomes progressively less efficient.
This is well documented in the recent high-resolution observations from the largest solar telescope DKIST \citep{2025ApJ...990L...3T}, where the authors show in their Figure 1 that the ribbon area stays practically the same during tens of minutes of the flare evolution, while the ribbon brightness fades significantly which we explain as a chromospheric temperature decrease.

\cite{Bicz_2025} analyzed a sample of stellar flares observed by TESS and ground-based photometric surveys, intending to estimate flaring areas and investigate their temporal evolution. Based on observations of seven new flares and eight previously published events \citep{Howard_2020} on M- and K-type main-sequence stars, the study showed that while flare areas remained approximately constant during most of the event (typically relative change varying by less than 30\%), the flare temperatures evolved significantly, often increasing by a factor of $\sim\!\!2.5$. These findings challenge the commonly used assumption of constant temperature (typically $10\,000\,$K) in flare energy estimations and suggest that the thermal evolution plays a more dominant role in shaping the observed flare light curves. The analysis of the events presented in \cite{Bicz_2025} also shows that the flare temperature may exceed $10\,000\,$K only for a brief fraction of the event duration (e.g., around 5\% of the total time), or may not reach this value at all.
Furthermore, flare areas estimated using the constant-temperature method (Method I) can differ significantly from results obtained with the more detailed analysis (Method IIb - see Fig. \ref{fig:comparison}), ranging from several times smaller to more than an order of magnitude larger. Consequently, Method I can overestimate flare energies by factors of several for some events, while systematically underestimating higher-energy flares by up to an order of magnitude. In contrast, Method IIb provides more reliable estimates of both flare area and energy across the full sample.

One of the most general, though admittedly imprecise, approaches to estimating flare energy is the assumption that the maximum kinetic temperature of the flare reaches approximately $T=10\,000\,$K. Based on this value, one can initially estimate the emitting area using the Method I. In contrast to the original approach, where the flare area is allowed to vary over time, we propose keeping the area fixed while allowing the temperature to evolve (Method II). As previously discussed, this assumption is more physically justified and forms the basis of the modification to the Method I that we propose.

Unlike Method~I, which assumes constant temperature and variable area, our approach fixes the flare area and determines time-dependent temperature evolution. This change is physically motivated by solar and stellar observations showing quasi-stable magnetic topology but a significant thermal variation.

The total emitted energy of a flare ($E_{\rm flare}$) can be estimated using the stellar luminosity ($L_{\rm star}$), flare amplitude, duration, and the assumed maximum flare temperature ($T^{\rm max}_{\rm flare}$). This estimation assumes that the white-light flare spectrum is approximately blackbody one and that the flare region is optically thick, fully blocking the underlying photospheric emission (see also \citealt{Heinzel2024}). The total observed luminosity ($L$) from both the star and the flare is:
	\begin{equation}
	    L = \left(A_{\rm star} - A_{\rm flare}\right)F_{\rm star} + A_{\rm flare}
	         F_{\rm flare}{\rm ,}
	\end{equation}
	where $F_{\rm flare}$, $A_{\rm flare}$, $F_{\rm star}$, and $A_{\rm star}$ represent the surface fluxes and areas of the flare and the star, respectively. The luminosity of the star is described by:
	\begin{equation}
	 	L_{\rm star} = A_{\rm star} F_{\rm star}.
	\end{equation}
	The normalized light curve of the star ($C$) is then given by:
	\begin{equation}
	C = \frac{L}{L_{\rm star}} = \frac{\left(A_{\rm star} - A_{\rm flare}\right)F_{\rm star} + A_{\rm flare} F_{\rm flare}}{A_{\rm star} F_{\rm star}}.
	 \end{equation}
	This formula can be rewritten to express the area of the flaring region as:
	\begin{equation}
		A_{\rm flare} = \frac{(C-1)A_{\rm star}F_{\rm star}}{F_{\rm flare} - F_{\rm star}}.
	\end{equation}
    Note that here we no longer assume that $A_{\rm flare} \ll A_{\rm star}$.
	Assuming the flaring area remains constant throughout the event and it is proportional to the maximum amplitude of the flare, the above relation becomes:
	\begin{equation}\label{eq:aflare}
		A_{\rm flare} = \frac{\mathrm{max}(C-1) A_{\rm star} \int\limits_{\lambda_1}^{\lambda_2} S_{\lambda}
        \mathcal{F_{\rm star}}
        d\lambda}{\int\limits_{\lambda_1}^{\lambda_2} S_{\lambda} B_\lambda(T^{\rm max}_{\rm flare}) d\lambda - \int\limits_{\lambda_1}^{\lambda_2} S_{\lambda}
        \mathcal{F_{\rm star}}
        d\lambda},
	\end{equation}

	\noindent where $\lambda$ is the wavelength, $\mathcal{F_{\rm star}}$ is the synthetic or observed spectrum of the quiescent star (usually replaced by the Planck function at stellar effective temperature), $B_\lambda(T)$ is the Planck function, $T_{\rm star}$ and $R_{\rm star}$ are the effective temperature and radius of the star, respectively, and $S_{\lambda}$ is the response function of the instrument that observed the flare. After estimating the flare area for the assumed maximum temperature of the flare $T^{\rm max}_{\rm flare}$, Eq.~\ref{eq:aflare} can be rewritten as:
	\begin{equation}\label{eq:integral}
        \int\limits_{\lambda_1}^{\lambda_2} S_{\lambda} B_\lambda(T_{\rm flare}) d\lambda = \left(\frac{(C-1)A_{\rm star}}{A_{\rm flare}} + 1\right) \int\limits_{\lambda_1}^{\lambda_2} S_{\lambda}
        \mathcal{F_{\rm star}}\,d\lambda.
    \end{equation}
	Solving this integral equation for $T_{\rm flare}$ provides the time evolution of the flare's temperature during the event, allowing for the estimation of the flare's bolometric luminosity ($L_{\rm flare}$) as:
	\begin{equation} \label{eq:lflare}
		L_{\rm flare}(t) = \sigma_{\rm SB} T_{\rm flare}^4(t) A_{\rm flare}
	\end{equation}
	Both $L_{\rm flare}$ and $T_{\rm flare}$ are now functions of time. The total bolometric energy of the flare ($E_{\rm flare}$) is obtained by integrating $L_{\rm flare}$ over the duration of the flare:
	\begin{equation}\label{eq:etot}
		E_{\rm flare} = \int\limits_{\rm flare} L_{\rm flare}(t) \, dt = A_{\rm flare}\,\sigma_{\rm SB}\int\limits_{\rm{flare}} T^4_{\rm flare }(t) dt\, .
	\end{equation}

Here it has to be stressed that the area computed from Eq. \ref{eq:aflare} is the projected area depending on the flare position on stellar disk. However,
the area in Eq. \ref{eq:etot} must be the de-projected one in order to get the
total energy emitted by the continuum into all outgoing directions. But because
we can not simply determine the flare position (except of some rare cases
like those reported in \citet{2021MNRAS.507.1723I} and \citet{Bicz_20242}), we consider here all studied flares
as being located in the disk center. Later on we discuss the uncertainties
introduced by this common assumption.

\begin{figure}[ht!]
    \begin{center}
        \includegraphics[width=0.5\textwidth]{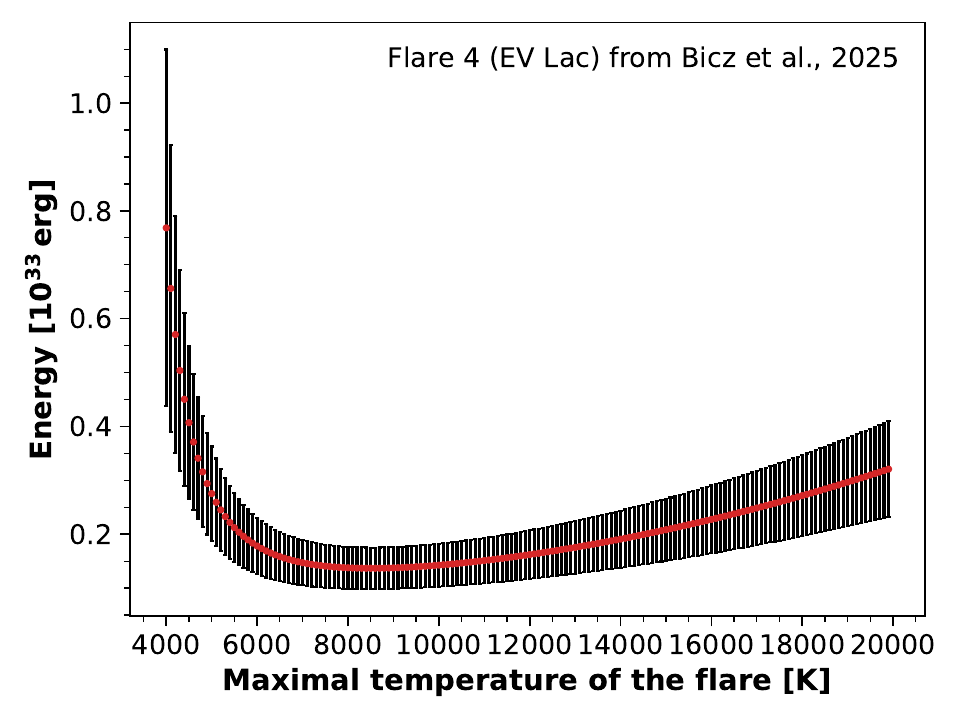}
    \end{center}
    \caption{The relationship between the flare energy estimated using our modified method IIa and the assumed peak flare temperature is illustrated for a representative event. The red dots show the energy values corresponding to different temperatures, while the black error bars indicate the associated uncertainties.}
    \label{fig:tenerpub}
\end{figure}

	The uncertainties in the stellar parameters, $R_{\rm star}$ and $T_{\rm star}$, taken from the TESS Input Catalog (TIC; \cite{Stassun2018}) and accessed via the MAST (Mikulski Archive for Space Telescopes) archive, are approximately $0.02\,R_\odot$ and $160\,\mathrm{K}$, respectively. Consequently, the uncertainty of the flare energy itself is about $\pm 28$\%. The area of the flare is influenced by both the flare's maximum amplitude and its assumed maximum temperature and
is primarily determined by the latter. We calculate it using Eq. \ref{eq:aflare}, which depends on the stellar parameters, the flare amplitude, and the flare temperature. Since the flare area is assumed to remain constant throughout the event, its value is fixed and determined based on the peak flare amplitude and temperature.
Importantly, the calculated area is relatively insensitive to moderate variations in this temperature. For example, a change of $\Delta T_{\rm flare} = 1000\,\mathrm{K}$ results in only a $~5\%$ difference in area, while $\Delta T_{\rm flare} = 2000\,\mathrm{K}$ leads to a $~10\%$ change.
Consequently, the total bolometric energy varies moderately with the peak temperature. This relationship is illustrated in Fig.~\ref{fig:tenerpub}, using a representative event from \citet{Bicz_2025}, where the estimated peak temperature is $10\,100 \pm 2300$\,K. The flare occurred on the well-studied M dwarf EV~Lac, which has an effective temperature of $T_{\mathrm{eff}} = 3315 \pm 152$\,K \citep{Paudel_2021};
note that in this paper we use the notation
$T_{\rm{eff}}$ as well as $T_{\rm{star}}$ for star's effective temperature.
Within the uncertainty range of the peak temperature, the resulting energy estimates differ by approximately 15\%. For higher temperatures (e.g., $16\,000$\,K), a change of $\Delta T_{\rm flare} = 2000$\,K leads to a similar $\sim\!16\%$ variation in energy. At lower temperatures, close to the stellar $T_{\mathrm{eff}}$, the uncertainties become more significant; however, the flare's contribution to the total energy in these cooler phases is typically negligible. Notably, the difference in estimated flare energy between assumed flare temperatures of $T_{\rm flare} = 6000$\,K and $T_{\rm flare} = 4000$\,K is about a factor of eight. Because the flare area is determined from the peak amplitude and temperature, changing the assumed peak temperature produces a characteristic curve of bolometric energy versus temperature with a broad minimum. This minimum effectively represents the lowest energy consistent with the black-body approximation (see Fig.~\ref{fig:tenerpub}); for temperatures on either side of it, the inferred energy rises again.

\subsection{Method IIb Using a Constant Flare Area and Assuming an Empirical Flare Temperature}\label{subsec:interpt}

\noindent The method for estimating flare energy under the assumption of a constant emitting area, as described in Subsection \ref{subs:constanta}, can be extended to incorporate a realistic, rather than arbitrarily assumed, maximum flare temperature. This modification is based on a more precise estimation of the peak flare temperature from observational data, which in turn allows for a more accurate determination of the total energy emitted during the event.

The overall procedure for calculating flare energy remains essentially the same as the one presented in Subsection \ref{subs:constanta} (Method IIa).
The main distinction concerns the way used to estimate the maximum flare temperature.
Instead of assuming a fixed, canonical value of $10\,000\,$K, our extended method employs a numerical grid introduced by \cite{Bicz_2025}, which enables the estimation of the maximum temperature based on the observed flare amplitude. Alternatively, if multi-band optical observations are available, an independent color-temperature method may be used to estimate the flare’s actual temperature.
In both approaches, the resulting time evolution of the flare temperature $T_{\rm{flare}}(t)$ can be derived consistently.

It should be emphasized, however, that this approach is currently applicable primarily to main-sequence stars of spectral types K4V and later.
This limitation is particularly relevant for data from the TESS satellite, where the host star’s effective temperature is known. By using this information, the method eliminates the need for the canonical assumption of a maximum flare temperature, potentially leading to more accurate energy estimates. Consequently, it allows for a more realistic characterization of flare energies, particularly when comparing flare intensities across different stellar types.

\begin{figure}[ht!]
    \begin{center}
        \includegraphics[width=0.5\textwidth]{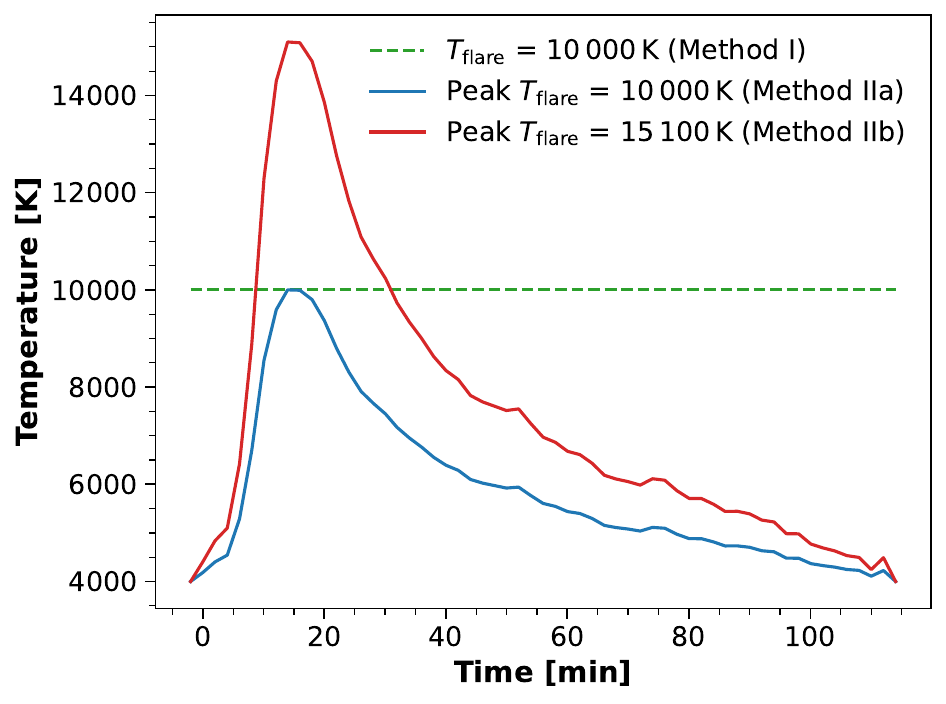}
    \end{center}
    \caption{Time evolution of the flare temperature for a superflare TIC 325178532 for three different methods. Note an asymptotic temperature decrease towards stellar $T_{\rm eff}$ for methods IIa and IIb.}
    \label{fig:temp}
\end{figure}

\section{Example of temperature evolution in a TESS superflare}

\noindent To assess the accuracy of our modified energy determination method, we applied it to a superflare observed by TESS on the star TIC 325178532, previously analyzed by \citet{Pietras22}. This star has an effective temperature $T_{\rm eff}=4000 \pm 120$ K and a radius of $R_{*}=0.9 \pm \, 0.1\, R_{\odot}$ (MAST\footnote{https://archive.stsci.edu/} catalog).

Temporal evolution of the flare temperature, illustrated in Fig.~\ref{fig:temp}, begins at the stellar effective temperature ($T_{\rm eff} = 4000$ K) prior to the impulsive rise phase. At the flare peak, the temperature reaches its maximum value, after which it gradually declines back to the pre-flare level. Throughout this evolution, the flare area is assumed to remain constant, corresponding to the value derived at the peak temperature.

\begin{table}[ht]
\centering
\caption{Comparison of flare energy estimates for TIC 325178532 using different methods.}
\label{tab:method_comparison}
\begin{tabular}{lccc}
\hline
Method & Temperature Assumption & Flare Area & Total Energy [erg] \\
\hline
Method~I  & $T = 10\,000\,K\ \mathrm{(constant)}$ & Variable: $4223 \pm 595 \,\mathrm{ppm}$ at the peak & $5.06 \times 10^{34}$ \\
Method~IIa & $T(t),\ \mathrm{peak} = 10\,000\,K$ & Fixed: $4439 \pm 658 \,\mathrm{ppm}$ & $4.33 \times 10^{34}$ \\
Method~IIb & $T(t),\ \mathrm{peak} = 15\,100\,K$ & Fixed: $1873 \pm 276 \,\mathrm{ppm}$ & $7.12 \times 10^{34}$ \\
\hline
\end{tabular}
\end{table}

Figure~2 and Table~\ref{tab:method_comparison} demonstrate that although Method~IIa and Method~I produce similar results when using the same peak temperature (10{,}000\,K), the inclusion of realistic peak temperature in Method~IIb significantly alters the outcome. For TIC 325178532, the flare energy increases by 40\% compared to Method~I. Across a large sample (50$\,$320 flares), Method~IIa yields lower energies than Method~I in $\sim$92\% of cases, while Method~IIb gives higher energies in $\sim$55\% of cases, with extreme discrepancies exceeding an order of magnitude. These statistics confirm that our method captures thermal evolution and peak temperature effects that strongly influence flare energetics.

To estimate the total flare energy, we compare three distinct methods, each based on different assumptions regarding flare temperature and emitting area:
\begin{enumerate}[labelindent=\parindent, leftmargin=*]
    \item[--] Method I --  Standard approach: Using the standard method introduced by \cite{Shibayama13}, which assumes a constant flare temperature of $T = 10\,000$ K and a blackbody emission spectrum, we calculate the total energy to be $E_{\rm tot} = 5.06 \times 10^{34} \,\mathrm{erg}$. This approach
    assumes a constant flare temperature throughout the event and does account for the area evolution. The resulting flare area at the peak is estimated to be $A_{\rm flare} = 4223 \pm 595 \,\mathrm{ppm}$.\footnote{Parts per million}
    \item[--] Method IIa -- Fixed peak $T_{\rm flare}$ with fixed area approach (Subsection \ref{subs:constanta}): In our implementation with the constant-area model, we assume the same peak temperature of $T = 10\,000\,$K but we allow the temperature value to vary over time. The flare area in this case is $A_{\rm flare} = 4439 \pm 658 \,\mathrm{ppm}$ which is comparable to the peak value in Method I. This yields a slightly lower total energy of $E_{\rm tot} = 4.33 \times 10^{34} \,\mathrm{erg}$. Despite accounting for the drop in temperature after the peak, the total energy differs only moderately from the Method I.
    Notably, the flare area in this model remains fixed, while the temperature is treated as variable throughout the flare duration.\
    \item[--] Method IIb -- Interpolated peak $T_{\rm flare}$ with fixed area approach (Subsection \ref{subsec:interpt}): In this approach, we apply the semi-empirical temperature–amplitude grid from \citet{Bicz_2025}, which allows us to estimate the peak flare temperature more realistically. For the K7V-type star TIC 325178532 and an observed flare amplitude of 0.08679, we derive a peak temperature of $T_{\rm peak} = 15\,100 \pm 2700 \,\mathrm{K}$, resulting in a flare area at the peak $A_{\rm flare} = 1873 \pm 276 \,\mathrm{ppm}$, which is more than a factor of two smaller than in the previous method.
    When the full time-dependent temperature evolution is taken into account, the total energy increases to
$E_{\rm tot} = 7.12 \times 10^{34} \,\mathrm{erg}$,
approximately 40\% higher than the Method I. This significant difference reflects the strong sensitivity of bolometric luminosity to temperature (see Eq.~\ref{eq:lflare} and conclusions contained in \cite{Bicz_2025}). Although the flare temperature exceeded 10\,000 K for less than 19\% of the flare duration, its influence on the total emitted energy was substantial.
\end{enumerate}

\section{TESS data analyzed with the new methods}

\noindent After validating our method on a single TESS superflare, the next step was to evaluate how the energy estimates from different approaches compare across a broad sample. To this end, we applied both the Method I and our revised technique—in two variants as Method IIa (Subsection \ref{subs:constanta}) and
Method IIb (Subsection \ref{subsec:interpt}), using either a fixed peak flare temperature of $10\,000\,\mathrm{K}$ or a temperature estimated from the semi-empirical grid — to a dataset comprising $50\,320$ flares observed on 268 of the most active stars from \citet{Pietras22}. We further extended the sample of flares on these stars by including data from TESS up to Sector 94. The sample consists of 268 stars with spectral types ranging from K4V to M7V, which corresponds to the validity range of the flare-temperature grid presented in \cite{Bicz_2025}.
For the photometry, we use the TESS PDCSAP (Pre-search Data Conditioned Simple Aperture Photometry) flux light-curve products provided by the MAST/SPOC (Science Processing Operations Center; \cite{Jenkins_2016}) pipeline. These light curves have undergone a co-trending procedure to remove systematic instrumental effects, as detailed by \cite{Smith_2012}. To ensure uniform and reliable flare analysis, we exclusively use the short-cadence PDCSAP time series and retain only data points with quality set to zero, thereby minimizing the influence of instrumental systematics and data anomalies.

\begin{figure}[!ht]
\begin{center}
\includegraphics[width=\textwidth]{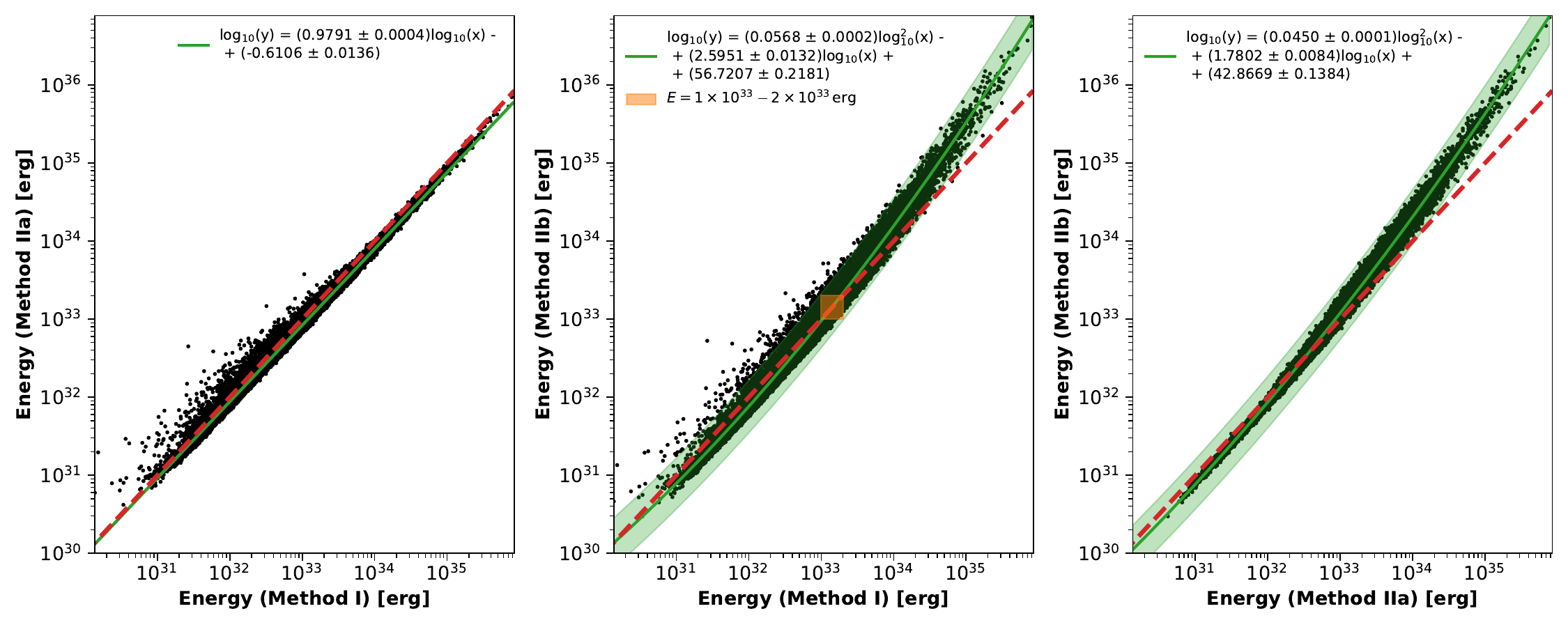}
\end{center}
\caption{Comparison of three methods for estimating the bolometric energy of flares. The panels show the energy estimated with Method I versus Method IIa (left), Method I versus Method IIb (middle), and Method IIa versus Method IIb (right). The green shaded areas indicate the uncertainty of the polynomial fits (linear for the left panel, quadratic for the middle and right). The dashed red lines represent a one-to-one correspondence. The orange square in the middle panel marks the energy range between $1\times 10^{33} - 2\times10^{33}\,$erg at which the quadratic fit intersects the one-to-one line. Note that the points are plotted as non-transparent to ensure visibility in the final scale; consequently, significant overlap in the panels may affect the visual impression of point density. While the distribution may visually appear balanced, the underlying data confirms that 92\% of cases yield lower energies for Method IIa compared to Method I.}
\label{fig:comparison}
\end{figure}

The resulting comparison is shown as a set of scatter plots in Fig.~\ref{fig:comparison}. When we compute flare energies using our method with a fixed peak flare temperature of $10\,000\,$K (Method IIa) and compare them to estimates from the Method I, we find that in approximately 92\% of cases, our method yields lower bolometric energy values. However, when the peak flare temperature is instead estimated using the Method IIb, about 55\% of the resulting energy estimates are higher than those obtained with the Method I, and 76\% gives higher estimates compared to our fixed peak flare temperature (Method IIa). Furthermore, when we use the interpolated peak temperatures from Method IIb, the resulting bolometric energies closely match those from the Method I, provided the flare temperature is around $12\,000 \pm 1\,400\,$K, corresponding to an energy range of approximately $1 \times 10^{33}\,$erg to $2 \times 10^{33}\,$erg. For peak temperatures below this range, the Method I leads toward higher estimates of flare energies, whereas for higher temperatures, it tends to underestimate them (the energy can be two times lower). For the updated distribution of flare temperatures see Fig. \ref{fig:histo}. In extreme cases, the discrepancy between the two methods can exceed an order of magnitude. This shows that in our sample the upper limit of flare energy is $\sim\!10^{37}\,$erg. It must be noted that the current grid is based on a small sample of 15 flares. Enlarging this sample to produce more robust peak temperature estimations is the subject of our ongoing research. To determine the empirical relation between energies (see Fig.~\ref{fig:comparison}), we perform a polynomial regression in log–log space. Specifically, we fit a first-order and second-order polynomial to the data. The $1\sigma$ uncertainties of each coefficient are then taken as the square root of the diagonal elements of the covariance matrix.

\section{Discussion and conclusions}

\noindent We have proposed a modified approach for estimating stellar flare energies that assumes a constant flaring area while allowing the flare temperature, interpreted as the kinetic temperature of the chromospheric condensation, to evolve over time. This method leverages the peak blackbody temperature to estimate the flare area, which then remains fixed throughout the whole event.

Our results clarify how earlier constant-temperature models lead to
considerable systematic errors: they neglected the thermal evolution of flares
and relied on arbitrary temperature assumptions. By incorporating time-dependent
temperature evolution and empirically constrained peak temperatures, our
approach greatly reduces these biases and provides more accurate flare energy
estimates. While differences relative to Method~I are modest when both
approaches assume a temperature of 10 000\,K, adopting realistic peak
temperatures can change the inferred energies of individual events by up to an
order of magnitude in large statistical samples. This improvement is crucial for
robust comparisons across flares and for assessing the effects of stellar
activity on planetary environments.

Historically, flare energies have often been estimated using simplified
methods such as Method~I \citep{Shibayama13}, which assumes a constant
temperature of 10 000~K and a time-varying emitting area inferred from the
light-curve amplitude. These assumptions have been widely applied in large-scale
statistical studies (e.g., with \textit{Kepler} and TESS data) to enable flare
comparisons. However, such simplifications can introduce systematic errors of up
to an order of magnitude in bolometric energy estimates.

This has important implications for solar–stellar comparisons, where flare energy is a key parameter in scaling relations (e.g., emission measure versus heating rate) and in evaluating whether stellar flares are scaled-up analogues of solar flares or represent fundamentally different physical phenomena. Both underestimation and overestimation of flare energies can obscure their true impact on stellar atmospheres and exoplanetary environments. Therefore, it is essential to employ flare energy estimation methods that reproduce the true energy release as accurately as possible.

As an illustrative example, the flare on Wolf~359 analyzed by \citet{2023ApJ...954...19P} was classified as solar X-class based on X-ray diagnostics (GOES-
equivalent class X3.7--X5.7), yet its bolometric energy inferred from TESS data
was only $\sim 1.1 \times 10^{31}$~erg, corresponding to an X-ray equivalent of
approximately M2.1. The loop geometry and plasma parameters closely resemble
those of strong solar flares, supporting the idea that flares share the same
underlying physical mechanism--provided that their energies are estimated
accurately.

Using older methods (constant temperature, Method~I) can bias energies for large stellar flares by factors of 2--10, and for superflares, errors can exceed an order of magnitude, shifting their position on emission-measure--energy diagrams and distorting statistical distributions of flare energies. For instance, for a TESS superflare, Method~I gives $5.06 \times 10^{34}$\,erg, while a realistic temperature model (Method~IIb) gives $7.12 \times 10^{34}$\,erg—a 40\% difference. In extreme cases, discrepancies exceed ten times. Adopting improved methods (time-dependent temperature, empirical peak temperature) ensures consistent placement of stellar flares on solar-derived diagrams and reliable extrapolation of solar conditions to stellar ones.

Here an obvious question may arise why we develop such a new method when it statistically results in energies not dramatically different form previous determinations. Our motivation was to suggest a method which is physically consistent with commonly accepted scenario of temperature variations during the flare evolution - flares start from a more-or-less quiescent pre-flare atmosphere, peak due to the significant heating and end-up again as quiet regions. But fortunately, this modified scenario statistically doesn't change
much the generally accepted picture of superflares.

 For superflares, we assume (as \cite{Shibayama13}) that the continuum spectrum
 is Planckian at a given kinetic temperature. This happens when the chromospheric
 condensation layer has so large densities that the Paschen recombination
 continuum is optically thick enough to ensure the thermalisation of the
 continuum radiation towards the the Planck
 function. However, as demonstrated by \cite{Heinzel2024}, at lower
 densities the synthetic spectrum lies below the level of the Planck function
 and thus becomes dependent on the column emission measure within the condensation,
 namely during the gradual phase of the flare. This behaviour may influence
 the total flare energy determination, but using only one-channel data like
 {\it Kepler} or TESS, one cannot obtain any information about the emission
 measure. Therefore, our energy determination here represents an upper limit
 of the superflare energy. But assuming that the major part of this energy
 is related to the impulsive phase with high densities, our estimate seems
 to be reasonable.

Determining the upper energy limit of a flare in the \cite{Heinzel2024} model is not straightforward and requires a detailed analysis of the physical properties of the chromospheric layer. A preliminary analysis indicates that continuum flare energy estimates are generally reliable only when the chromosphere becomes optically thick. At lower column emission measures, the chromosphere remains optically thin, causing the derived energy to depend strongly on the column emission measure as well as on the stellar and flare temperatures (e.g. \cite{Heinzel2024}). Under these conditions, the peak flare energy may not coincide with regions of highest density or largest column emission measure for a given flare peak temperature and stellar surface temperature. Understanding this effect requires further investigation. It should be emphasized, that for weaker flares, however, one would need the multi-band photometry or broadband spectral observations to calculate the changes in the area and the temperature of the flare. Note that our method, as well as the original one of \cite{Shibayama13}, is proposed to make a statistical estimate of the superflare energies based on extended samples of space data. Such statistics can hardly be obtained using ground-based photometry or even spectroscopy. For such extended samples of space data we are unable to get information about the flare location on the stellar disk and we thus assume the position in the disk center, in the same way as
\cite{Shibayama13}. This then introduces another source of non-negligible
uncertainties: assuming that the positional cosine $\mu$=0.5, the de-projected
flare area will be twice larger and thus also the energy.

\begin{figure}[ht!]
    \centering
    \includegraphics[width=0.65\textwidth]{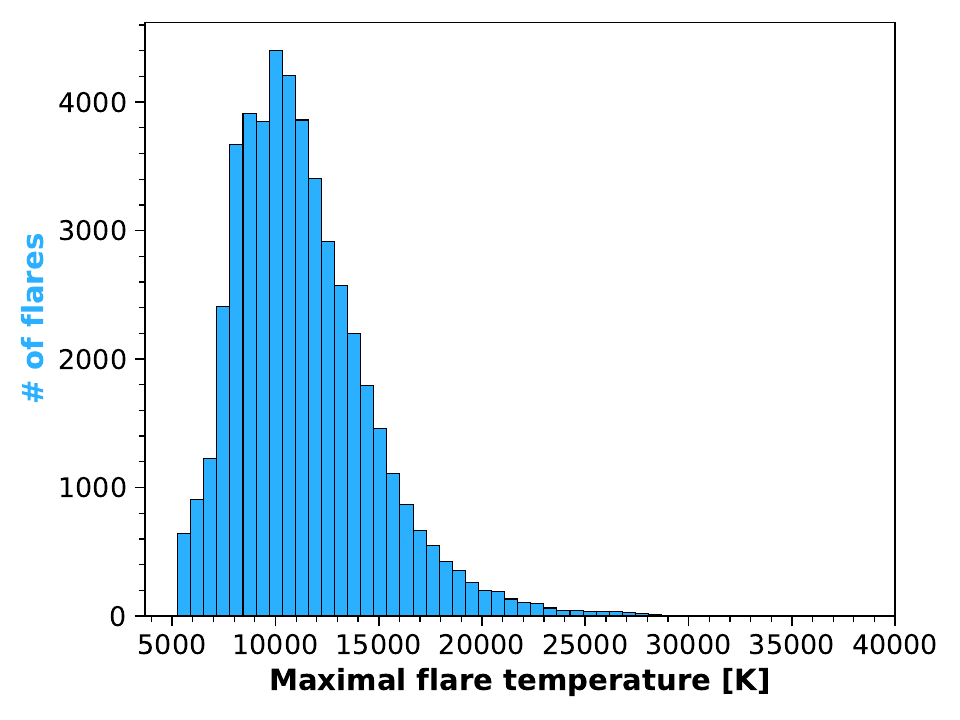}
    \caption{Distribution of the peak temperatures of 50$\,$320 flares on 268 stars, from \cite{Pietras22} and \cite{Bicz_2025}, estimated using our semi-empirical grid.}
    \label{fig:histo}
\end{figure}

The results of the extended analysis presented in this work, which includes observations from additional TESS sectors (up to sector 94) for a sample of 268 of the most active stars, clearly confirm that in over 50\% of the analyzed flares, the peak temperature reaches approximately $11\,000 \pm 2\,000\,$K (see Fig. \ref{fig:histo}).
We derive the characteristic flare temperature from the distribution of maximal temperatures in our sample. A lognormal probability density function (PDF) is fitted to the data using maximum likelihood estimation, and the mode of the fitted distribution (the temperature at which the PDF is maximized) is taken as the representative peak value.
Uncertainty is quantified by evaluating the fitted PDF within a symmetric interval of $\pm 2\,000\,$K around the mode. This interval captures the region where the majority of the probability density is concentrated, and in practice, it contains more than half of the analyzed flares. Thus, the characteristic peak temperature is 11$\,$000$\,$K with an uncertainty of $\pm 2\,000\,$K, reflecting the spread of maximal flare temperatures around the distribution’s maximum.
This finding significantly expands upon our previous analysis \citep{Bicz_2025} and suggests that adopting a higher peak temperature in bolometric energy estimation methods may be more physically justified.
Therefore, we propose using a peak temperature of $11\,000\,$K as a more
representative value in Method IIa,
instead of the traditionally assumed $10\,000\,$K used in Method I.

The application of our new method to large-scale sky surveys, combined with a revision of the average peak flare temperature from the canonical 10\,000 K to 11\,000 K, leads to a reduction in the errors associated with estimating the total energy of the analyzed stellar flares.
However, it should be emphasized that the most accurate energy estimates are obtained when the peak flare temperature is well constrained, either through multi-band optical observations which are currently not available on a large scale, or by using the amplitude–energy relation grids, assuming a known stellar effective temperature for main-sequence stars of spectral types K4V and later.

\mbox{}\hfill\\
\noindent This work was partially supported by the program "Excellence Initiative - Research University" for the years 2020 - 2026 for the University of Wroc\l{}aw, project No. BPIDUB.4610.96.2021.KG. P.H. also acknowledges support from the project RVO:67985815 of the Astronomical Institute of the Czech Academy of Sciences.
The computations were performed using resources provided by Wroc\l{}aw Networking and Supercomputing Centre (https://wcss.pl), computational grant number 569. This paper includes data collected by the TESS mission. NASA’s Science Mission Directorate provides funding for the TESS mission.
We thank the anonymous referee for helpful comments and suggestions.

\bibliographystyle{aasjournal}
\bibliography{bibliography.bib}

\end{document}